\documentclass[sigconf]{acmart}
\renewcommand\footnotetextcopyrightpermission[1]{}
\def\BibTeX{{\rm B\kern-.05em{\sc i\kern-.025em b}\kern-.08emT\kern-.1667em\lower.7ex\hbox{E}\kern-.125emX}}

\setcopyright{none}

\usepackage{xspace}
\usepackage{listings}
\usepackage{tikz}
\usepackage[noend]{algpseudocode}
\usepackage{tabularx,booktabs}
\usepackage[english]{babel}
\usepackage{pdfpages}
\usepackage{adjustbox}
\usepackage[bookmarks=false]{hyperref}
\let\oldReturn\Return
\renewcommand{\Return}{\State\oldReturn}

\newcommand{\hide}[1]{}

\newcommand{\code}[1]{\texttt{#1}\xspace}

\definecolor{pblue}{rgb}{0.13,0.13,1}
\definecolor{pgreen}{rgb}{0,0.5,0}
\definecolor{pred}{rgb}{0.9,0,0}
\definecolor{pgrey}{rgb}{0.46,0.45,0.48}
\definecolor{ncs}{rgb}{0.0, 0.53, 0.74}

\algnewcommand\algorithmicswitch{\textbf{switch}}
\algnewcommand\algorithmiccase{\textbf{case}}
\algnewcommand\algorithmicassert{\texttt{assert}}
\algnewcommand\Assert[1]{\State \algorithmicassert(#1)}%
\algdef{SE}[SWITCH]{Switch}{EndSwitch}[1]{\algorithmicswitch\ #1\ \algorithmicdo}{\algorithmicend\ \algorithmicswitch}%
\algdef{SE}[CASE]{Case}{EndCase}[1]{\algorithmiccase\ #1}{\algorithmicend\ \algorithmiccase}%
\algtext*{EndSwitch}%
\algtext*{EndCase}%

\newcolumntype{R}[2]{%
	>{\adjustbox{angle=#1,lap=\width-(#2)}\bgroup}%
	l%
	<{\egroup}%
}
\newcommand*\rot[2]{\multicolumn{1}{R{#1}{#2}}}

\newcommand\eatpunct[1]{}

\begin{document}
\pagenumbering{gobble}

\title[Security Implications Of Compiler Optimizations]{Security Implications Of Compiler Optimizations On Cryptography -- A Review}

\author{Ashwin Prasad Shivarpatna Venkatesh}
\email{ashwin@campus.uni-paderborn.de}
\affiliation{
  \institution{Paderborn University}
  \country{Germany}
}

\author{Aditya Bhat Handadi}
\email{abh@campus.uni-paderborn.de}
\affiliation{
	\institution{Paderborn University}
	\country{Germany}
}

\author{Martin Mory}
\email{martin.mory@upb.de}
\affiliation{
  \institution{Paderborn University}
  \country{Germany}
}

\thispagestyle{empty}

\begin{center}	

	\colorbox{ncs}{
		\begin{minipage}{17cm}
			\begin{minipage}{.68\textwidth}
  			{\color{white}
				 \vspace{1.7cm}
				{\hspace{1.1em}\fontsize{30}{60}\selectfont\textbf{Technical Report}} \\ [20pt]
				 \vspace{.2cm}{\hspace{1.1em}\huge\textbf{Paderborn University}} \\
				 \vspace{.2cm}{\hspace{1.1em}\huge\textbf{tr-ri-19-358}} \\
				 \vspace{.2cm}{\hspace{1.1em}\huge\textbf{\today}} 
				\vspace{1.5cm}
			}
		\end{minipage}%
		\begin{minipage}{.32\textwidth}
  				\includegraphics[width=4.8cm]{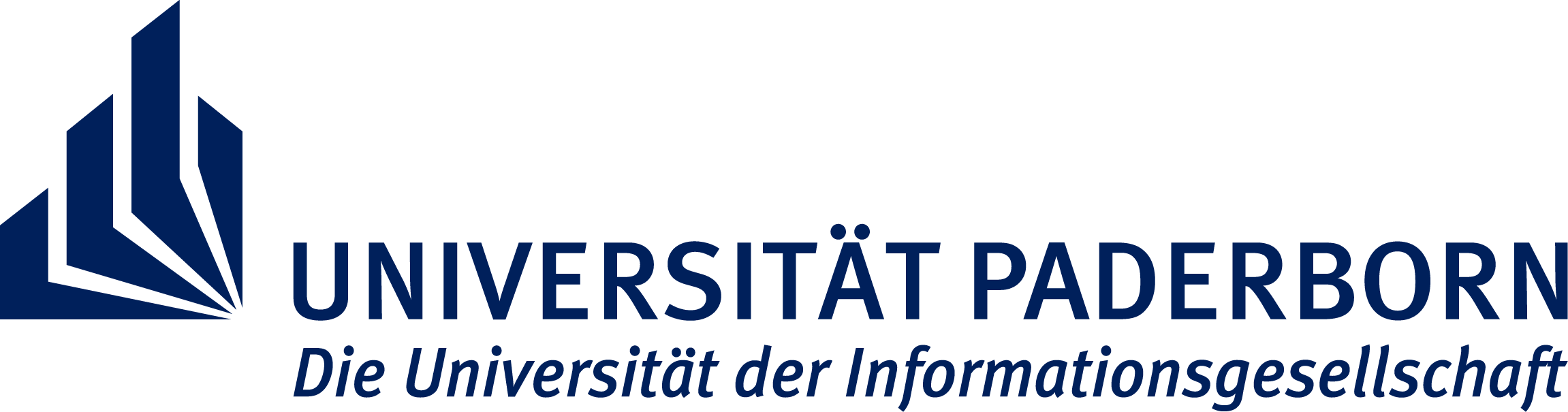} \\ [20pt]
  				\includegraphics[width=4.8cm]{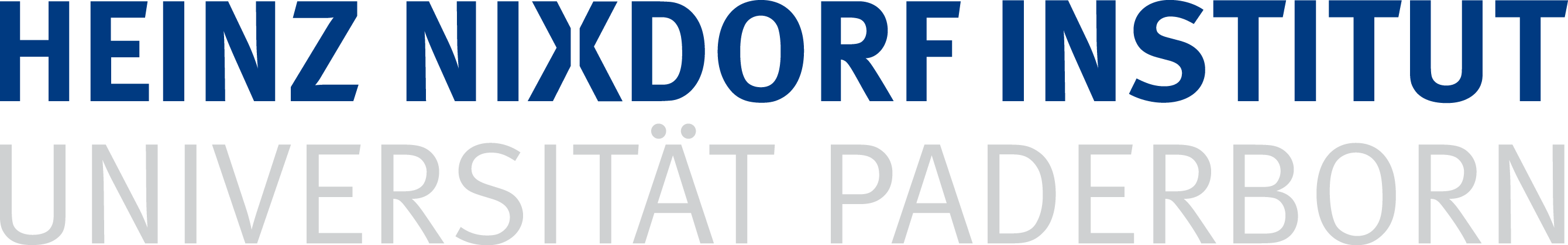}
		\end{minipage}%
		\end{minipage}
	}
		
	 \vspace{1.5cm}
	{\fontsize{20}{60}\selectfont\textbf{The Security Implications Of Compiler}} \\[10pt]
	{\fontsize{20}{60}\selectfont\textbf{Optimizations On Cryptography -- A Review}} 
	 
	\vspace{2cm}
	\includegraphics[height=4.5cm]{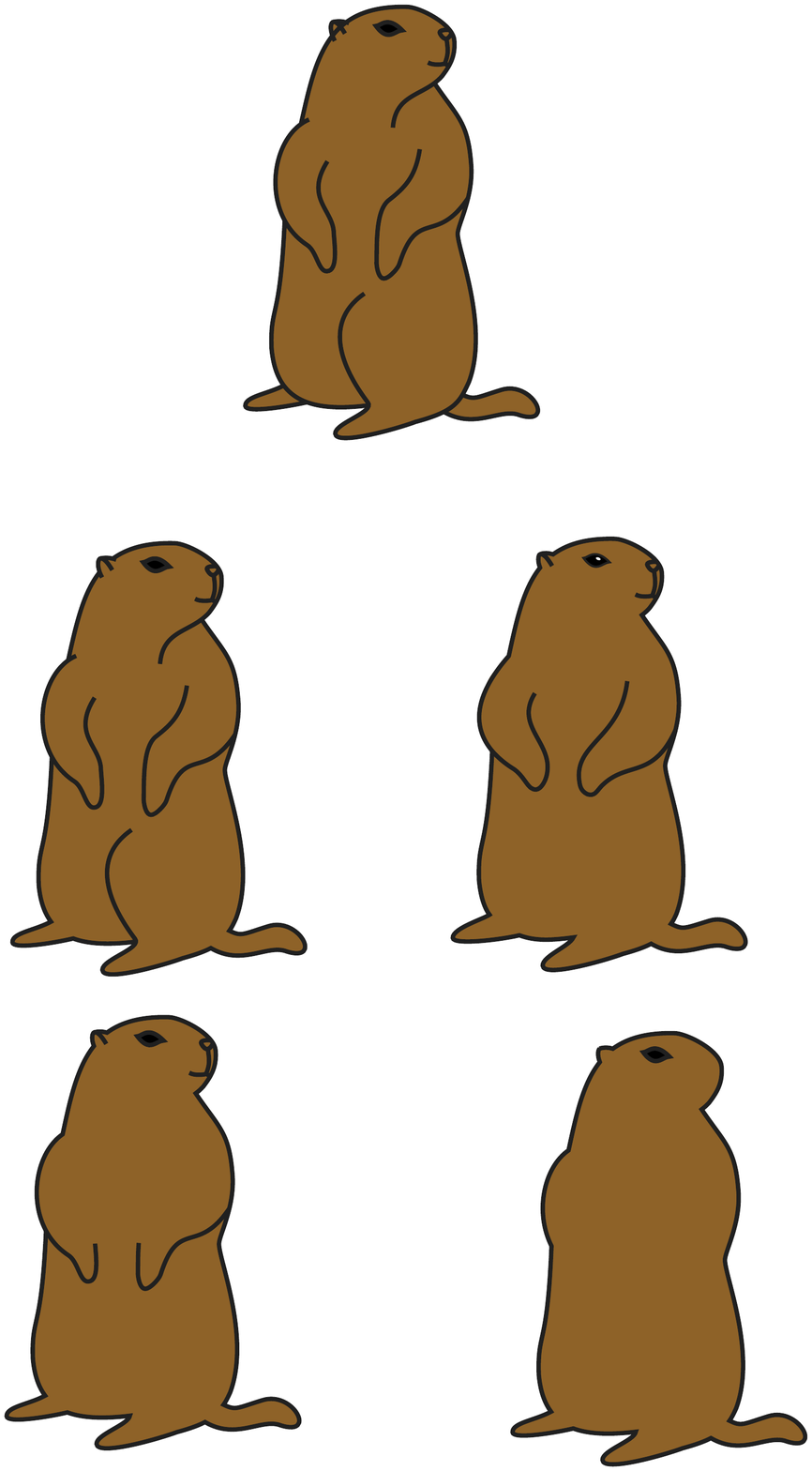}
	\vspace{2cm}
	 
	\begin{minipage}{17cm}
  		{\Large
			\textbf{Authors:} \\ [5pt]
			{Ashwin Prasad Shivarpatna Venkatesh (Paderborn University)} \\
			{Aditya Bhat Handadi (Paderborn University)} \\
			{Martin Mory (Paderborn University)} \\[5pt]
		}
	\end{minipage}	
	
	\vspace{.2cm}
	\colorbox{ncs}{
		\begin{minipage}{17cm}
			{ \hspace{17cm}
			\vspace{.5cm}}
		\end{minipage}
	}

\end{center}
\newpage

\clearpage
\pagenumbering{arabic}
\begin{abstract}

When implementing secure software, developers must ensure certain requirements, such as the erasure of secret data after its use and execution in real time. Such requirements are not explicitly captured by the C language and could potentially be violated by compiler optimizations. As a result, developers typically use indirect methods to hide their code's semantics from the compiler and avoid unwanted optimizations. However, such workarounds are not permanent solutions, as increasingly efficient compiler optimization causes code that was considered secure in the past now vulnerable. 

This paper is a literature review of (1) the security complications caused by compiler optimizations, (2) approaches used by developers to mitigate optimization problems, and (3) recent academic efforts towards enabling security engineers to communicate implicit security requirements to the compiler. In addition, we present a short study of six cryptographic libraries and how they approach the issue of ensuring security requirements.
With this paper, we highlight the need for software developers and compiler designers to work together in order to design efficient systems for writing secure software.

\end{abstract}

%
%


\maketitle

\sloppy

\section{Introduction}

Making software secure requires the assurance of its consistency when executed. Developers must thus ensure that certain properties are respected, although some cannot be explicitly conveyed to the compiler. For example, cryptographic algorithms must run in constant time to be secure against timing attacks. However, there is currently no way of communicating this intention to the compiler. As a result, the programmer might use indirect techniques in order to achieve the intended goals by controlling the side effects of the target language and its compiler. 

On the other hand, when designing compile time optimizations, compiler designers are more concerned about the defined standards of the language than those properties required by software developers. With more efficient compilers being produced every day, techniques used by the programmers might be broken by more advanced compiler optimizations in the future.

This creates a counter-productive race between software developers and compiler designers. We advocate for both parties to work in conjunction, in order to achieve a more simple and efficient compiling/development system that can yield more secure software. In this paper, we discuss the compiler optimizations that break security properties, and recent work that proposes and demonstrates modifications to the compiler systems to allow developers to specify desirable program properties.

This paper is structured as follows. We first introduce background information about compiler optimization in Section~\ref{Background}. We then present an overview of the problems caused by the requirement gap between software developers and compiler designers in Section~\ref{Problem}. Section~\ref{Approach}, details various approaches used to mitigate the side effects created by compiler optimization and further explain the implementation of some of these approaches. In Section~\ref{casestudies}, we investigate open-source cryptographic libraries and the techniques they use to control the side effects. Finally, we present the related work in Section~\ref{related}, the future work in Section~\ref{future}, and conclude in Section~\ref{conclusion}.

\section{Background}
\label{Background}

Compilers translate program instructions from one programming language to another. In the case of C / C++, the high-level program code written by the developer is translated into machine-level instructions which can be executed on a processor. The translation operation also optimizes the program, for example by detecting and removing unused variables, or dead code. \emph{Compiler optimization} is the process of improving the performance of the translated code without changing its functionality. Primary attributes such as the execution time, the code size, or the compile time are typical candidates for optimization. However, some optimization techniques can interfere with the security properties encoded in the source code, which are then lost after the translation.

In this section, we present a few compiler optimization techniques which are relevant for the remainder of this paper. We focus on optimizations for C, which have been observed to alter the expected security properties of a program~\cite{dsilvaCorrectnessSecurityGapCompiler2015}, and present the compiler flags used to enable or disable them. Unfortunately, none of those optimizations can be explicitly controlled in the compiler yet, apart from disabling them.

\paragraph{Dead Store Elimination (DSE)}
\label{optidse}
DSE is an optimization used to reduce execution time and memory usage. It finds memory store operations that are either not used or overwritten and removes those instructions. This is an issue when developers design their code to scrub parts of the memory which were used for sensitive data storage. DSE may remove the scrubbing instructions, and thus expose sensitive data in memory. This is illustrated in Listings~\ref{dsec} and ~\ref{dsea}, where the key value in \code{key} is explicitly overwritten in the C code, but is optimized out in the assembly code after compilation. 
The mainstream C-compiler has no options or flags to explicitly disable certain instructions from being optimized. Even though the C programming language standard \textit{C11} includes a solution to that issue: \code{memset\_s}, a secured version of \code{memset}, there is no standard-compliant implementation yet~\cite{yangDeadStoreElimination}.
In recent research, Yang et al.~\cite{yangDeadStoreElimination} and Simon et al.~\cite{simonWhatYouGet2018} have addressed this problem by implementing support for secure dead store elimination, which we discuss in detail in the Section~\ref{implsecreterasure}.

Compiler flag: \textbf{\code{-fdse} :} Perform DSE.

\begin{figure}[t]
\begin{lstlisting}[language=c,caption=Dead Store Elimination -- C Code,label=dsec]
int dummy(int x){
  int y = x+1;
  return y;
}

int secret_function(){
  int key = 0xDEADBEEF;
  int y = dummy(key);
  key = 0x00; // <---- Missing in assembly code
  return y;
}
\end{lstlisting}

\begin{lstlisting}[caption=Dead Store Elimination -- Assembly Code,label=dsea]
; clang 3.9 -O1  -m32 -march=i386
dummy(int):
mov     eax, dword ptr [esp + 4]
inc     eax
ret

secret_function():
sub     esp, 12
mov     dword ptr [esp], -559038737
call    dummy(int)
add     esp, 12 ; <---- Missing mov 0 (Optimized out by DSE)
ret
\end{lstlisting}
\end{figure}

\paragraph{Link Time Optimization (LTO)} LTO is an inter-procedural optimization technique which analyzes all of the compilation units of a program and optimizes it as a single module. This expands the scope of the optimization to a global view. Some workarounds programmers use in order to ensure program security work on the scope of small modules, for example by depending on variables defined in other modules, so that the compiler cannot resolve their values and optimize them out. Compiling the program on a global scale using LTO defeats such workarounds.

Compiler flag: \textbf{\code{-flto} :} Enable LTO.

\paragraph{Dead Code Elimination (DCE)} DCE is a compiler optimization that removes \emph{dead code}: code that is never executed or does not change the outcome of the program. This results in a smaller code base and reduces the run time of a program by removing unused operations. However, just like DSE, DCE might remove implicit operations required by the code developer, for instance, a call to zero a buffer.

Compiler flag: \textbf{\code{-fdce} } Perform DCE.

\paragraph{Tail-Call Optimization (TCO)}
\label{tcoopti}
TCO is a technique that optimizes stack accesses during function calls. A \emph{tail call} is a function call located at the end of the caller function. TCO replaces such calls with jump instructions, so that the stack frame of the calling function is reused, as opposed to creating a new stack frame for the callee. This optimizes memory usage. However, this can lead to the removal of sensitive function calls, as described by Simon et al.~\cite{simonWhatYouGet2018}.

Compiler flag: \textbf{\code{-foptimize-sibling-calls} } Optimize sibling and tail recursive calls.

\section{Complications Caused by Compiler Optimizations}
\label{Problem}

In this section, we discuss the main problems caused by the translation issues on compilation. We first detail the existing mechanisms offered by the standard C compiler to address some of those issues, which are called \emph{portability issues} in the C standard~\cite{c1999}. We then expand on the two most popular security requirements not supported by the C compiler: the \emph{implicit Invariants}.

\subsection{Portability Issues}
\label{expinv}

The C standard~\cite{c1999} introduces the following three types of behavior that encode assumptions the compiler makes.

\paragraph{Unspecified Behavior (USB)} is defined as the ``Use of an unspecified value or other behavior where this International Standard provides two or more possibilities and imposes no further requirements on which is chosen in any instance. \textit{Example:} Order in which the arguments to a function is evaluated." (Section 3.3.4 from~\cite{c1999}).

\paragraph{Implementation-Defined Behavior (IDB)} is the ``Unspecified behavior where each implementation documents how the choice is made \textit{Example:} the size of types" (Section 3.4.1 from~\cite{c1999}).

\paragraph{Undefined Behavior (UB)} is the ``Behavior, upon use of a nonportable or erroneous program construct or of erroneous data, for which this International Standard imposes no requirements \textit{Example:} behavior on integer overflow." (Section 3.4.3 from~\cite{c1999})

As these properties are explicitly captured by the standard, developers can use flags and options to control some of the compile-time behavior, such as the ones described in Section~\ref{Background}. However, this system is not enough to capture more complex requirements, as described in the following section.

\subsection{Implicit Invariants} 
\label{implinvar}

Along with the documented explicit behavior (USB, IDB, and UB), software developers must also encode more indirect behavior that is critical to ensure security properties: the \emph{implicit Invariants}, two of which we present below.

\paragraph{Constant-Time Selection}
\label{ctselection}
Many security operations such as password checking or cryptographic operations should ideally always execute with the same duration. If not, the systems open side-channels and are vulnerable to branch prediction, pipeline stalling, and timing attacks. Figures~\ref{ctsc} and~\ref{ctsa} illustrate the issue through a branching example. In the C code, whether the if-branch or the else-branch is executed, the result would be the same to an external observer: since one return operation is run, it would take the same time, so the observer cannot determine which branch the program took. However, in the assembly code, the else-branch has one more instruction: the jump instruction. So despite the C-code being able to obfuscate paths, an observer would be able to tell which branch was used based on how long the program runs.
Compiler optimizations can thwart the developer's efforts in obfuscating the path taken by the program.
As of now, there is no option nor compiler flag that a developer can use to communicate the implicit requirement of constant-time selection to the compiler. As a result, developers resort to writing complex logic to outsmart the compiler and prevent it from optimizing out certain operations. We discuss this in more detail in Section~\ref{customconstsel}.

\begin{figure}[t]
\begin{lstlisting}[language=c,caption=Constant-Time Selection Problem -- C Code,label=ctsc]
int conditional_select(bool b, int x, int y){
  if(b){
    return x;
  } else {
    return y;
  }
}
\end{lstlisting}

\begin{lstlisting}[caption=Constant-Time Selection Problem -- Assembly Code,label=ctsa]
; clang 3.9 -O3  -m32 -march=i386
conditional_select(bool, int, int):
mov     al, byte ptr [esp + 4]
test    al, al
jne     .LBB0_1  ; <--- JUMP
lea     eax, [esp + 12]
mov     eax, dword ptr [eax]
ret
.LBB0_1:
lea     eax, [esp + 8]
mov     eax, dword ptr [eax]
ret
\end{lstlisting}
\end{figure}

\paragraph{Secret Erasure}
\label{secerasure}
A major concern when handling sensitive data is to perform a memory scrub to erase it from memory after its use. An example is shown in Listings~\ref{dsec} and~\ref{dsea}, where a sensitive key is erased from the RAM. A common technique to erase memory is to use the \code{memset} function. However, the compiler can assume that isolated calls to \code{memset} are not useful, since the memory that is written to by the \code{memset} will not be read again. To improve performance, a DSE optimization typically removes such calls, leaving sensitive data exposed at runtime. 
The current solution in C11 is the \code{memset\_s} function, which is guaranteed to be never removed by optimization. However, its use is still not widely supported~\cite{simonWhatYouGet2018}.

\subsection{Desirable Compiler Properties For Cryptography}
\label{usecases}

In past research, D'Silva et al.~\cite{dsilvaCorrectnessSecurityGapCompiler2015} have looked at the particular case of cryptography, and have identified basic implicit invariants required to support developers. Table~\ref{genusecases} details the main three invariants, and why they should be supported by compilers. Those invariants revolve around the protection against side-channels at runtime. A detailed discussion on the attacks is available in D'Silva et al.'s paper~\cite{dsilvaCorrectnessSecurityGapCompiler2015}, Section III.C.

\begin{table}[t]
	\caption{Implicit invariants for cryptography}
	\begin{tabular}{p{3cm}p{4.5cm}}
		\toprule 
		\rule[-1ex]{0pt}{3.5ex} \textbf{Purpose} & \textbf{Use case} \\ 
		\midrule
		\rule[-1ex]{0pt}{3.5ex}Constant Side Effects & Avoid side-channel attacks \\ 
		\rule[-1ex]{0pt}{3.5ex}Constant Control Flow &  Protection against timing attacks \\ 
		\rule[-1ex]{0pt}{3.5ex}Constant Memory Access &  Prevent cache-based side-channel attacks\\ 
		\bottomrule 		
	\end{tabular} 
	\label{genusecases}
\end{table}

Table~\ref{gentech} describes the main measures implemented by developers in their C code to avoid the attacks mentioned in Table~\ref{genusecases}. Those measures aim at obfuscating the program's footprint, involving secret erasure---as seen in the previous section---, noise addition, and bit splitting and splicing. However, compiler optimizations can render such techniques useless, as illustrated in Listings~\ref{dsec}--~\ref{ctsa}. This motivates the need for more control on compiler optimizations for the developer, or for more cooperation between compiler designers and software developers.

\begin{table}[t]
	\caption{Techniques used by developers to preserve security properties}
	\begin{tabular}{l p{5.5cm}}
		\toprule 
		\rule[-1ex]{0pt}{3.5ex} \textbf{Technique} & \textbf{Explanation} \\ 
		\midrule 
		\rule[-1ex]{0pt}{3.5ex} Noise Addition & Adding arbitrary noise to confuse side-channel attacks \\ 
		\rule[-1ex]{0pt}{3.5ex} Bit Splitting & Scattering data across memory to make reconstruction harder for attacks on RAM dumps\\ 
		\rule[-1ex]{0pt}{3.5ex} Bit Splicing &  Splitting variables into bits and utilizing bit-wise operators to mitigate timing attacks\\ 
		\rule[-1ex]{0pt}{3.5ex} Secret Erasure & Scrubbing sensitive data from RAM after usage\\ 
		\bottomrule 
	\end{tabular} 
	
	\label{gentech}
\end{table}

\section{Practical Mitigations of Insecure Compiler Optimizations}
\label{Approach}

In this section, we present an overview of existing approaches used by developers or developed by researchers to circumvent compiler optimization problems.

\subsection{Custom Functions For Constant-Time Selection}
\label{customconstsel}

To ensure constant-time selection, developers typically write convoluted implementations in order to control the side-effects introduced by compiler optimizations. One such technique is to avoid the usage of \code{bool} and use custom cure to compare integers instead. Alternatively, some libraries provide functions to support constant-time integer comparison, such as such as OpenSSL\footnote{\url{https://www.openssl.org/}}, which contains 37 of them~\cite{simonWhatYouGet2018}.

In Listing 4 of their paper~\cite{simonWhatYouGet2018}, Simon et al. list four different versions of such indirect approaches for constant-time selection and they show that the implementation is not consistent across different versions of GCC at different optimization levels (we refer the reader to Tables 1 and 2 of Simon et al.'s paper~\cite{simonWhatYouGet2018}). 

\subsection{Custom Functions For Stack Erasure}
\label{cffse}
In past research, Yang et al.~\cite{yangDeadStoreElimination} provide a universal scrubbing function as an easy-to-use C file, which combines the best approaches found in the real-world open source projects. Their library file also allows developers to specify their preference with respect to the scrubbing order. The scrubbing techniques they support include (1) platform-provided scrubbing functions (e.g., SecureZeroMemory and \code{memset\_s}), (2) the memory barrier technique, (3) the volatile data pointer technique and, (4) the volatile function pointer technique.

\subsection{Disabling Optimization}

As an extreme measure, disabling the optimization can prevent unexpected behavior after compilation. GCC provides predefined optimization levels such as \code{-O1, -O2, -O3, ...}~\cite{gccdocumentationoptions} which enables the corresponding supported optimizations. When no optimization level is specified, the compiler does not perform any optimization. However, disabling optimization is often discouraged since it leads to excessive performance overhead~\cite{dsilvaCorrectnessSecurityGapCompiler2015}.

\subsection{Hiding Semantics}

Many scrubbing techniques consist in hiding the semantics of their scrubbing operations from the compiler. The rationale is that if the compiler fails to recognize that an operation is clearing memory, it will not remove it. \emph{Separate compilation} is one such technique where a scrubbing operation is implemented in a separate compilation unit. The compiler fails to remove calls to the scrubbing function because it does not know that it is equivalent to \code{memset}. However, separate compilation is not reliable when link time optimization (LTO) is enabled, since all compilation units are merged into one, giving the compiler a global view of the whole program~\cite{yangDeadStoreElimination}. Thus, to ensure the success of this technique, the developer needs to have control over how the program is compiled, and disable LTO.

Another popular technique for hiding a scrubbing operation from the compiler is to call the memory scrubbing function through a \emph{Volatile function pointer}~\cite{yangDeadStoreElimination}. \code{OPENSSL\_cleanse} of OpenSSL 1.0.2, shown in Listing~\ref{opensslcleanse1}, uses this technique. The effect of declaring \code{memset\_func} as volatile means that the compiler must read its value from memory every time it uses it, because the value may have changed. The compiler does not know the value of \code{memset\_func} at compile time, since it cannot recognize the call to \code{memset} and remove it. Yang et al.~\cite{yangDeadStoreElimination} have confirmed that this technique works effectively on GCC, Clang and Microsoft Visual C.

\begin{figure}[t]
\begin{lstlisting}[caption = OpenSSL volatile function pointer, label = opensslcleanse1]
typedef void *(*memset_t)(void *,int,size_t);
static volatile memset_t memset_func = &memset;
void OPENSSL_cleanse(void *ptr, size_t len){
	memset_func(ptr, 0, len);
}
\end{lstlisting}
\end{figure}

\subsection{Forcing Memory Writes}

An alternative technique is to force the compiler to add the store operation without concealing it. Two major techniques are used to this end. \emph{Complicated computation} makes use of a function that reads and writes garbage data from a global variable to the memory that needs to be scrubbed, thus filling it with garbage. Listing~\ref{opensslcleanse2} shows an example of scrubbing using complicated computation from OpenSSL, prior to version 1.02. The function \code{OPENSSL\_cleanse} uses the global variable \code{cleanse\_ctr}, which provides varying garbage data. Since global variables can be accessed from anywhere in the program, it is difficult for the compiler to determine whether a function such as \code{OPENSSL\_cleanse} is should be optimized out, without running an inter-procedural analysis on the entire program~\cite{yangDeadStoreElimination}. This kind of analysis is considered too costly for compilers to perform.

\begin{figure}[t]
\begin{lstlisting}[caption = OpenSSL\_cleanse, label = opensslcleanse2]
unsigned char cleanse_ctr = 0;
void OPENSSL_cleanse(void *ptr, size_t len) {
	unsigned char *p = ptr;
	size_t loop = len, ctr = cleanse_ctr;

	if (ptr == NULL) return;

	while (loop--) {
		*(p++) = (unsigned char)ctr;
		ctr += (17 + ((size_t)p & 0xF;
	}
	p = memchr(ptr, (unsigned char)ctr, len);

	if (p) ctr += (63 + (size_t)p);
		cleanse_ctr = (unsigned char)ctr;
}
\end{lstlisting}
\end{figure}

The other technique, \emph{memory barrier}, is supported both by GCC and Clang. Through an inline assembly statement, a simple memory argument specifies the compiler that the statement may access and modify the memory, thereby forcing the compiler to retain the stores instead of terming them as dead~\cite{yangDeadStoreElimination}. A more reliable way to define a memory barrier is illustrated by Linux's \code{memzero\_explicit}, as shown in Listing~\ref{memzeroexplicit}. The difference is the \code{r(ptr)} argument that makes the pointer to the scrubbed memory visible to the assembly code, and stops the scrubbing store from being removed.

\begin{figure}
\begin{lstlisting}[caption = Linux memzero\_explicit, label = memzeroexplicit]
#define barrier_data(ptr) \
__asm__ __volatile__("": :"r"(ptr) :"memory")

void memzero_explicit(void *s, size_t count) {
	memset(s, 0, count);
	barrier_data(s);
}
\end{lstlisting}
\end{figure}

\subsection{Platform-Supplied Functions}

The most convenient way to ensure a memory scrub is to use a dedicated function. Windows provides a \code{SecureZeroMemory} implementation which is guaranteed to be secure from optimization. Microsoft Visual Studio supports this initiative by never optimizing out a call to \code{SecureZeroMemory}. Although it is found to be effective, it is currently only available on Windows. 

Another alternative introduced by the ANSI C standard C11 is \code{memset\_s} function which is declared as follows:\\
\code{errno\_t memset\_s(void* s,rsize\_t smax,int c,rsize\_t n)} \\
Similar to \code{memset}, the \code{memset\_s} function sets a number of the bytes starting at address \code{s} to the byte value \code{c}. The number of bytes written is the minimum of \code{smax} or \code{n}. The two buffer sizes guard against overflows. Despite those guards, \code{memset\_s} can be misused, for example, by setting \code{smax} or \code{n} to 0. Thus, the function would fail to clear the buffer while preventing a buffer overflow. Although \code{memset\_s} seems like an ideal solution, its implementation is slow, which we attribute to the following reasons. Firstly, \code{memset\_s} is just part of the optional Appendix K of C11, and was not a required part of the standard. Secondly, C11 treats all functions in Appendix K as a single unit, forcing the library to implement all of the functions defined in the annex. Finally, the poor adoption and perceived flaws of \code{memset\_s} have led to calls for its removal from the standard.

\section{Communicating Security Requirements to Compilers}
\label{Implementationneval}

In this section, we discuss the recent work of Yang et al.~\cite{yangDeadStoreElimination} and Simon et al.~\cite{simonWhatYouGet2018} that allows software developers to communicate their security requirements to the compiler. They add explicit support for the \emph{constant-time selection} and \emph{secret erasure} implicit invariants to Clang\footnote{\url{https://clang.llvm.org/}}/LLVM\footnote{\url{https://llvm.org/}}.

Clang/LLVM is a compiler framework that consists of three components: the frontend, the optimizer, and the backend. The frontend translates the source code to an intermediate representation called LLVM IR. The optimizer is responsible for optimization transformations on LLVM IR, such as DSE or CSE. The backend translates the LLVM IR intermediate representation to the target machine language and performs target-specific optimizations.

\subsection{Constant-Time Selection}
\label{implconstant}

Simon et al.~\cite{simonWhatYouGet2018} add the following built-in function into the Clang/LLVM framework to support constant-time selection based on a boolean condition. Their implementation is available online~\cite{ctcclang}.
\code{\_\_builtin\_ct\_choose(bool condition, Type x, Type y)}		 
The built-in function returns \code{x} if \code{condition} is true, and \code{y} otherwise, taking constant time for both cases. The \code{condition} bool can be a comparison operator such as \code{==}, \code{!=}, etc. The integers \code{x} and \code{y} must be of the same integer type in the compiler front-end, otherwise, the function yields an error.

The authors add support to the x86\_64 backend by compiling the function into assembly code that uses the conditional move instruction \code{CMOV} instead of branches. This instruction was shown to ensure constant time selection~\cite{simonWhatYouGet2018}) after other optimizations are applied. For other backends, the function is compiled into a \code{XOR} instruction which has a generally higher probability to be constant.  

The authors evaluated \code{\_\_builtin\_ct\_choose} using two cryptographic implementations: OpenSSL's X25519 and a self-written constant-time RSA exponentiation using the Montgomery ladder~\cite{joye2002montgomery}. They used a tool called Dudect~\cite{reparazDudeMyCode2017} to empirically verify constant-timeliness by using millions of different inputs.
Measuring the CPU cycle overhead, they observed that the built-in solution has less than 1\% overhead for X25519 and is 4\% faster with RSA exponentiation.	

The usage of a single function such as \code{\_\_builtin\_ct\_choose} has the potential to improve code readability and usability for the developers, avoiding complicated workarounds and also guaranteeing constant-time selection for future versions of the compiler. However, the \code{\_\_builtin\_ct\_choose} built-in function only circumvents LLVM optimizations but not backend optimizations.


\subsection{Secret Erasure}
\label{implsecreterasure}
Simon et al.~\cite{simonWhatYouGet2018} add secret erasure support to the Clang/LLVM framework by using function annotations. Before detailing their three approaches to achieve reliable erasure, we first introduce background knowledge about the assumptions they make, and the underlying infrastructure of their work.

\subsubsection{Background}

\paragraph{CPU, OS and ABI} When compiling a C program, a large number of libraries and supporting applications, such as platform code, libc, runtime loader/linker, and Virtual Dynamic Shared Object (VDSO) are typically required. All of those third-party programs need to be recompiled after making the changes to the compiler framework for erasure to be effective over the entire system. In particular, signal handling in the Linux kernel can be complex when the signal is handled by storing the CPU state on the stack before stopping the execution. If a program is located in a sensitive memory block, sensitive data could be leaked to the stack, so it must be erased too.

\paragraph{Compiler and linker} Similarly to the libraries, the runtime library also needs to be recompiled. The compiler optimizes implementations of commonly used functions, sometimes inlining them for performance's sake. The effect is a change in the usage of registers, which should be handled with regards to security.
Many other compiler features and optimizations also alter the stack, for instance, tail-call optimization, defer-pop optimization, shrink-wrapping optimization, function multi-versioning, or static linker stubs.

\paragraph{The programmer} The developer's role is critical to ensure proper erasure in their program. Certain non-returning functions need to be handled separately, because, unlike other functions, the stack cannot be erased before returning. The developer should ideally avoid calling such functions in sensitive code. Variable-sized stacks should also be avoided. 

\subsubsection{The Function-Based Approach (FB)}
The first method for ensuring secret erasure is the \emph{function-based} approach (FB)~\cite{simonWhatYouGet2018}. It performs stack and register erasure for every sensitive function and its callees before it returns, using an annotated function. Non-returning functions are not supported by the approach. Tail-call optimizations must be disabled for sensitive functions, since they would make returning functions non-returning, but the authors decided to globally disable it in their demonstrations. Two variants are implemented, one with a signal handler (FB with SH) and another without (FB no SH). 

The authors evaluated their approach on OpenSSL, using mbedTLS to contain the instrumentation, and MiBench to measure the overhead.
Programs using FB with SH show to run 3.39$\times$ slower than without, while FB no SH programs are 1.86$\times$ slower. They also observed that FB solutions are generally not optimal since callee functions erase the same stack area repetitively.

\subsubsection{The Stack-Based Approach (SB)} 
The second approach--the \emph{stack-based} approach (SB)--instruments all of a program's functions to keep track of the run-time stack, using the global variable \code{\_\_GlobalStackValue}. When an annotated function returns, the stack is cleared using the offset value that is calculated based on the global variable. The implementation is available online, in this file \code{X86ZeroStackPass.cpp} \cite{x86stackimple}. The authors decided to implement this functionality in the function epilogue, which is executed just before the \code{ret}, because most registers are not live and will not be spilled to the stack. This also means that this approach only works for returning functions. Hence, tail call optimizations must be disabled with SB as well.
Simon et al. implemented two variants of SB: one with \emph{bulk register zeroing}--which zeros all registers at once in annotated functions (SB with BRZ), and one on which registers are erased in every function individually (SB no BRZ).

The approach is verified on OpenSSL, with mbedTLS for the instrumentation. The overhead measured with SB by this solution is significantly better than FB, with only 2\% overhead for SB no BRZ and 1\% for SB with BRZ. SB is faster because it operates with the registers to track the stack size, while with FB, the memory is erased for every function.
Note that instrumenting the musl-libc library increases its size by 6.6\% and 4.9\% for \textit{SB no BRZ} and \textit{SB with BRZ} respectively.

\subsubsection{The Call-Graph Based Approach (CGB)}
The last approach by Simon et al. is the call-graph based approach (CGB). It determines the maximum stack usage of a sensitive function at compilation time to eliminate the need for callee instrumentation. The maximum stack usage information is used as a heuristic to achieve a smaller footprint. The approach creates a call-graph of the program, which is used to identify the registers that would be written to, and the maximum stack usage for all annotated functions, thereby removing the need to instrument every function.
This approach supports tail call optimization is supported by this approach. Software developers are simply required to annotate the function pointers. A limitation of CGB is that call-graphs cannot handle infinite recursive functions, so a maximum depth should be specified.

The only overhead caused by this approach is the actual erasure which cannot be avoided. CGB is the fastest and most compact solution of the three approaches. A detailed comparison of the three approaches can be found in Table 3 of Simon et al.'s paper~\cite{simonWhatYouGet2018}.

\subsection{Secure Memset Implementation}
\label{scrubbingsafefunction}

Sometimes, developers do not endorse a specific compiler and rely on more manual techniques to prevent scrubbing operations from being optimized. For this use case, Yang et al.~\cite{yangDeadStoreElimination} have developed a standalone scrubbing function \code{secure\_memzero}. This function integrates effective scrubbing techniques such as \textit{platform supplied functions}, \textit{volatile function pointers} and the \textit{memory barrier technique}, as described in Section~\ref{cffse}.
The implementation is done in a header file \code{secure\_memzero.h}, which can be included in a C/C++ source file. An order of preference for the different techniques can be specified using macros. By default, the implementation is carried out in the order mentioned above.

\subsection{Inhibiting Scrubbing DSE}
\label{inhibitingdse}
Yang et al.~\cite{yangDeadStoreElimination} also implement a \emph{scrubbing-safe dead store elimination} option in Clang 3.9.0. It prevents DSE optimization from removing scrubbing operations by identifying the potential scrubbing operations beforehand. 

A store instruction can be marked as a memory scrubbing operation if:
\begin{itemize}
	\item The value being stored is a constant.
	\item The number of bytes is constant.
	\item The store will be eliminated because it goes out of scope without being read.
\end{itemize}

In their implementation, the authors retain all dead stores satisfying the conditions above regardless of whether they are sensitive or not. This leads to false positives when some non-sensitive dead stores are preserved. The authors argue that this implementation does not require any changes to the underlying source code, and only relies on the scrubbing techniques discussed in Section~\ref{cffse}.
The authors also note that with the help of developers, who can mark the actual sensitive data, the false positives rate can be improved. Developer work is reduced from using complicated workarounds to annotating store instructions.

On the SPEC 2006 benchmark, the implementation shows an overhead of around 1\%. By completely disabling DSE from clang, the performance overhead remains under 2\%, except for the 403.gcc benchmark, where it reached 5\%.

\section{Case Studies}
\label{casestudies}

In this section, we investigate large open-source cryptographic libraries and projects that use techniques to implicitly control compiler optimizations, or that implement and propose such techniques for developer usage. A summary of our findings is found in Section~\ref{summaryCaseStudies}.

\subsection{BearSSL}

BearSSL\footnote{\url{https://www.bearssl.org/}} is an implementation of the Transport Layer Security (TLS) and Secure Sockets Layer (SSL) protocols (RFC 5246) written in C, with a focus on portability, size and flexibility. BearSSL follows the guidelines of ``The Cryptography Coding Standard (CCS)" for the implementation of constant-time operations in C. Their implementation can be found online, in the file \code{src/inner.h}~\cite{Bearsslinner}. Constant-time conditional copy is achieved using the \code{br\_ccopy()} \cite{Bearssccopy} function.

In addition, constant-time AES and DES implementations in BearSSL use bit slicing techniques. Constant-time RSA and Elliptic Curves are enabled through a custom constant-time implementation of big integers.

BearSSL does not provide a secure memory wipe function but suggests using memset and an example code for stack erasure~\cite{Bearssmemory}.

\subsection{Monocypher}

Monocypher\footnote{\url{https://monocypher.org/}} is a lightweight, auditable cryptographic library written in portable C~\cite{monocyphergit}. It provides custom secure comparison functions such as \code{crypto\_verify16}, \code{crypto\_verify32}, and \code{crypto\_verify64}~\cite{monocypherc}.

Monocypher provides a secure memory wipe function \code{crypto\_wipe}~\cite{monocypherc} which uses the volatile pointer technique.

\subsection{Libsodium}
Sodium\footnote{\url{https://libsodium.gitbook.io/doc/}} is a cryptographic library which is a portable, cross-compilable, installable, and packageable fork of NaCl\footnote{\url{https://nacl.cr.yp.to/}}.
It provides a \code{sodium\_memzero}~\cite{libsodiumutils} function that first uses platform provided functions and then weak linkage techniques~\cite{yangDeadStoreElimination} to ensure secret erasure. Note that this may be insecure if compiled with LTO.

The \code{sodium\_memcmp} function~\cite{libsodiumutils} of Sodium ensures constant-time comparison. It also uses the weak symbols technique~\cite{yangDeadStoreElimination} or volatile pointers~\cite{yangDeadStoreElimination}, depending on what is available on the platform.

\subsection{Libgcrypt}
Libgcrypt\footnote{\url{https://www.gnupg.org/software/libgcrypt/index.html}} is another cryptographic library based on GnuPG. Libgcrypt implements a \code{wipememory} secure erasure based on the volatile data pointer technique~\cite{yangDeadStoreElimination}. However, with a recent commit~\cite{libgcryptcommit} the Libgcrypt authors encourage developers to favor the platform function \code{explicit\_bzero}~\cite{yangDeadStoreElimination} instead.

Libgcrypt also provides a custom function to perform constant-time comparison of two buffers: \code{buf\_eq\_const}~\cite{libgcryptbuf}.

\subsection{Crypto++}

Crypto++ is a C++ class implementing cryptographic algorithms. It defines \code{SecureWipeBuffer}~\cite{cryptoppmisc}, which scrubs memory by using custom assembly and the volatile data pointer technique~\cite{yangDeadStoreElimination}. Crypto++ also uses the \code{SecBlock} class, which provides a secure storage that is wiped when the block is destroyed.

In addition, Crypto++ contains the class \code{VerifyBufsEqual}~\cite{cryptoppmisccpp}, a constant-time comparison function utilizing bitwise operators.

\subsection{OpenSSL}
OpenSSL\footnote{\url{https://www.openssl.org/}} is a toolkit for the SSL/TLS protocols and a general-purpose cryptography library. It uses \code{OPENSSL\_cleanse}~\cite{opensslcleance} to scrub memory, defaulting to its own assembly implementations unless the \code{no-asm} flag is specified at configuration. Version 1.0.2 and above uses the volatile function pointer technique for calls to \code{memset}~\cite{yangDeadStoreElimination}.

Constant-time comparison of integers is supported in OpenSSL by implementing 37 different functions~\cite{simonWhatYouGet2018}.

\subsection{Summary}
\label{summaryCaseStudies}
Table~\ref{securetable} presents an overview of the secure erasure techniques used in the open-source projects and libraries presented in the previous sections. Each column represents a particular memory-erasure technique. \emph{Platform} represents secure scrubbing operations provided by the underlying platform, e.g. Windows' \code{SecureZeroMemory}, BSD's \code{explicit\_bzero}, or C11's \code{memset\_s}. \emph{Asm} is inline assembly code, and \emph{Volatile} includes two possible techniques: \emph{volatile data pointer} and \emph{volatile function pointer}. Those two techniques rely on the fact that volatile-qualified types are defined in the standard as having ``unknown side effects", thus they are not directly optimized by the compiler. \emph{Comp.} includes techniques which use complex computation to force the compiler to scrub memory. With the \emph{WL (Weak Linkage)} technique, the developer defines \emph{weak definitions}, a way of informing the compiler of future replacement at link time. Finally, \emph{memset} denotes the usage of the default memset function. Those six techniques are described in detail by Yang et al.~\cite{yangDeadStoreElimination}. Table~\ref{securetable} marks whether a project uses a certain technique, and if it does, it provides a prioritization number representing which technique is applied first, second, etc.

We see that different projects use a variety of techniques to reach the common goals of constant-time execution and secure memory erasure, some of which are custom-made and must be used with certain requirements. They help developers code with security requirements in mind, however, they only treat the symptoms of the problem. In addition to existing solutions, we advocate for developers and compiler designers to reach a mutual consensus and design a system that allows developers to have more control of the optimization options into the compiler.

\begin{table}[t]
	\centering
	\caption{Priorities of memory erasure techniques in open-source projects. A smaller number shows that the technique is preferred over a larger number. ``-'' denotes that the technique is not implemented in the project.}

\begin{tabular}{r|ccccccc}
	&
	\rot{45}{1em}{\textbf{Platform}} & 
	\rot{45}{1em}{\textbf{Asm}} & 
	\rot{45}{1em}{\textbf{Volatile}} & 
	\rot{45}{1em}{\textbf{Comp.}} & 
	\rot{45}{1em}{\textbf{WL}} & 
	\rot{45}{1em}{\textbf{memset}}	\\
	\midrule
\rule[-1ex]{0pt}{3.5ex}		Bearssl & - & - & - & - & - & - \\ 
\rule[-1ex]{0pt}{3.5ex}		 Monocypher & - & - & 1 & - & - & - \\ 
\rule[-1ex]{0pt}{3.5ex}		Libsodium & 1 & 3 & 4 & - & 2 & - \\ 
\rule[-1ex]{0pt}{3.5ex}		Libgcrypt & 1 & - & 2 & - & - & - \\ 
\rule[-1ex]{0pt}{3.5ex}		Crypto++ & - & 1 & 2 & - & - & - \\ 
	\rule[-1ex]{0pt}{3.5ex}		OpenSSL & - & 1 & 2 & 3 & - & - \\ 
	\bottomrule 
\end{tabular}
	\label{securetable}	
\end{table}

\section{Related Work}
\label{related}

Cauligi et al.~\cite{cauligiFaCTFlexibleConstantTime2017} argue that C is not suitable for both fast and readable code with cryptographic properties due to high-level constructs introducing timing vulnerabilities. They explore an alternate solution to use a domain specific language and a compiler that enable programmers to express implicit security properties, and produce timing-attack free code.

Abate et al.~\cite{abateWhenGoodComponents2018} study compartmentalized components and suggest how individual components should be protected from others even if they become compromised due to undefined behavior. The authors formally define a dynamic compromise of compartmentalized components and establish a criterion for secure compilation chain. This work highlights how one compromised component can potentially leak cryptographic keys from other components.

Reparaz et al.~\cite{reparazDudeMyCode2017} implement a minimal tool that verifies if a program runs in constant time on a particular platform, without needing to model the hardware. It relies on statistical analysis rather than static analysis.

Wang et al.~\cite{wangDifferentialApproachUndefined2016} explain how modern compilers exploit undefined behavior specifications of C/C++ to perform aggressive compiler optimization which introduces unpredictable outcomes. This details a formal and practical approach to find undefined behavior bugs by introducing a static checker called ``Stack" that identifies such bugs.

D'Silva et al.~\cite{dsilvaCorrectnessSecurityGapCompiler2015} detail the gap between the state of a program and the state of a machine. The authors propose accurate machine models to reason about the impact of compiler optimization on security, and also recommend future directions for research.

\section{Discussion and Future Work}
\label{future}
Despite being used in large software projects, the different approaches discussed in Section~\ref{Approach} are still not direct solutions to the compiler optimization problem. As a result, they run into limitations such as being weak to LTO, for example. While techniques that address the side effects are workarounds, the compiler-based approaches are a step in the right direction, addressing the main issue directly by introducing changes in the compiler itself. However, such solutions are still only academic, since adoption by the community is a difficult thing to achieve. 

In addition, compiler-based solutions also run into other limitations, as mentioned in Section~\ref{Implementationneval}. 
\begin{itemize}
\item Experimental results show a varying performance overhead associated with the compiler-based solutions.
\item The presence of false positives eaves window for further research into the topic for more accurate results.
\item The approaches discussed are limited to a single compiler framework, which would lead to poor adoption by the community.
\item Security requirements are largely undocumented in the C standard guide, which means that compilers which are following official guidelines do not need to support implicit security requirements.
\end{itemize} 

We argue that compiler development should consider developer intentions and requirements. Yang et al.~\cite{yangDeadStoreElimination} and Simon et al.~\cite{simonWhatYouGet2018} both advocate for compilers with explicit support for secure memory wiping and constant-time support as a starting point in the right direction. Going a step further, we suggest that research efforts should focus on how desirable properties can be made universal by adding support from the compiler framework. In the future, hardware based support for secure systems should also be explored, especially in situations where embedded systems are to be deployed in an open-world scenario and are prone to attacks. 

In parallel to research, we see value in spreading the word in industry about the importance of software security requirements in compiler optimization. Working groups can encourage the adoption of such requirements is existing standards, to bridge the gap between compiler designers and software developers.

\section{Conclusion}
\label{conclusion}

In this paper, we discussed how developer approaches to writing secure software often require to write convoluted code and design indirect solutions that outsmart compilers. We discussed existing solutions used in industry, and developed in current research, and argue that instead of using convoluted workarounds, researchers and practitioners should address the main problem directly, and work towards support for security considerations directly in the compiler framework. There is a large open area of collaborative research between software engineering and compiler design in how to make performant, multi-platform, compilers that support security requirements with respect to compiler optimization. Moreover, while the current state-of-the-art demands further research, a bigger challenge is the adoption of current studies and security requirements into mainstream compilers.

\begin{acks}
This research was conducted as part of the Secure Systems Engineering seminar at Paderborn University, organized by Eric Bodden and Lisa Nguyen Quang Do. It was partially funded by the Heinz Nixdorf Foundation.
\end{acks}

\bibliographystyle{ACM-Reference-Format}
\bibliography{paper}

\end{document}